\documentclass[reprint,prb,aps,superscriptaddress,showpacs]{revtex4-1}
\usepackage{graphicx}
\begin{document}
\title{Spin-wave coupling to electromagnetic cavity fields in dysposium ferrite}

\author{M.\ Bia\l{}ek}
\email{marcin.bialek@epfl.ch}
\affiliation{Institute of Physics, \'Ecole Polytechnique F\'ed\'erale de Lausanne (EPFL), 1015 Lausanne, Switzerland}
\author{A.\ Magrez}
\affiliation{Institute of Physics, \'Ecole Polytechnique F\'ed\'erale de Lausanne (EPFL), 1015 Lausanne, Switzerland}
\author{J.-Ph.\ Ansermet}
\affiliation{Institute of Physics, \'Ecole Polytechnique F\'ed\'erale de Lausanne (EPFL), 1015 Lausanne, Switzerland}

\date{\today}

\begin{abstract}
Coupling of spin-waves with electromagnetic cavity field is demonstrated in an antiferromagnet, dysprosium ferrite (DyFeO$_3$).
By measuring transmission at 0.2\textendash 0.35~THz and sweeping sample temperature, magnon-photon coupling signatures were found at crossings of spin-wave resonances with Fabry-P\'erot cavity modes formed in samples.
The obtained spectra are explained in terms of classical electrodynamics and a microscopic model. 
\end{abstract}


\maketitle
Coupling of matter and electromagnetic radiation \cite{Mills74} is a topic of great interest in solid state physics research because of their hybrid quantum nature. \cite{Khitrova06} In the THz range, phonon-polaritons are a well-know example of a light-matter coupling. Recently, polaritons in the THz region were shown with intersubband transitions,\cite{Anappara09, Todorov10} cyclotron resonance \cite{Bayer17} and plasmons \cite{Yu18} in two-dimensional electron gases, as well as with intermolecular transitions in organic materials. \cite{Damari19} In these systems, a strong-coupling regime can be achieved when losses are smaller than the exchange rate between light and matter, \cite{Paravicini-Bagliani17} giving rise to the vacuum Rabi splitting.  
Polaritons are composite particles, which are studied in basic research on quantum optics,\cite{Khitrova06} and can be considered for use in quantum computing and quantum memories.\cite{Wasenberg09, Schuster10, Kubo10, Huebl13, Kockum19}
The coupling of electromagnetic cavity-modes to magnons was researched intensively in ferromagnets at GHz frequencies, \cite{Huebl13, Yao15, Castel17} meeting with expectations of energy-efficient spintronic devices. \cite{Matsukura15} Coupling of magnons with superconducting qubits was also investigated. \cite{Tabuchi15, Lachance-Quirione17} The Purcell enhancement and the vacuum Rabi splitting were demonstrated in ferromagnetic materials.\cite{Zhang14, Yabuchi14, Bai15, Maier-Flaig16}

It is interesting to investigate magnon-photon coupling in antiferromagnetic materials \cite{Mills74} in view of their high-frequency spin dynamics,\cite{Kampfrath11, Jungwirth16, Baierl16, Nemec18, Schlauderer19} comparing to that of ferromagnets. As this phenomenon is readily taken into account by classical electrodynamics, it was accounted for in the analysis of optical investigations of antiferromagnetic materials, for instance in Ref.\ \cite{Balbashov85, Kozlov93, Mikhaylovskiy14, Mikhaylovskiy15NC, Mikhaylovskiy15PRB}. Some experimental reports focused on characterization of interaction of antiferromagnetic magnons with photons in FeF$_2$ \cite{Sanders78, Jensen97}, NiO \cite{Li11}, TmFeO$_3$ \cite{Grishunin18} and in ErFeO$_3$. \cite{Li18, Sivarajah19} 
Most of these reports use classical electrodynamics. A microscopic picture was developed in Ref.\ \onlinecite{Li18}.
Here, we confront the classical electrodynamic model with predictions of a microscopic model by estimating, quantitatively from our data, the strength of the interaction of antiferromagnetic spin-waves with electromagnetic cavity modes in dysprosium ferrite DyFeO$_3$ (DFO). 

Dysprosium ferrite is an orthogonally-distorted perovskite. 
It shows antiferromagnetic ordering below a N\'eel temperature of about 640~K \cite{White82}. Dzyaloshinskii-Moriya leads to a weak ferromagnetism
In DFO, the spin canting allows two antiferromagnetic resonance modes to be excited: the quasi-ferromagnetic (qFMR) and the quasi-antiferromagnetic (qAFMR),\cite{Zvezdin79, White82, Balbashov85, Koshizuka88, Kozlov93} which are excited by the magnetic component of radiation.\cite{Reid15} 

Our measurements were performed above 400~K, where both resonances, monotonously soften with rising temperature. 
We used polycrystaline samples for their isotropic properties, as large anisotropic properties in crystals might hinder the coupling. \cite{Bialek19} We have created disk-shaped samples of 1.0 and 0.6 mm thicknesses, thus having different cavity-mode spectra. 
A sample was placed in a furnace, which allowed to control temperature up to 700~K.
By using quasioptical methods, we measured transmission with our continuous-wave THz spectrometer based on frequency extenders to a vector network analyzer (VNA). \cite{Caspers16} 
This complex signal $S_{21}(f,T,H)$ is a function of frequency $f$, temperature $T$ and magnetic field $H$. We report its power amplitude in dB units and its phase in degrees. 
In order to extract a signal related to magnetic resonances, we measured transmission through a sample at different temperatures $T$. 
We obtained temperature-differential spectra by subtracting averaged spectra measured at subsequent temperatures:
\begin{equation}
\frac{\partial S_{21}}{\partial T} = \Delta T^{-1}(S_{21}(f,T+\Delta T,0) - S_{21}(f,T,0)),
\end{equation}
with $\Delta T =1$~K. We used this technique to measure spin-wave resonances in bismuth ferrite at high \cite{Bialek18} and low temperatures.\cite{Bialek19} 


In classical electrodynamics, the dispersion of electromagnetic radiation in a material with a resonance is modified. This leads to creation of two polariton states split by a frequency representing coupling of the resonance with the electromagnetic radiation.\cite{Mills74}
In DFO, we model magnetic susceptibility $\mu_c(f,T,H)$ using lorentzian distributions
\begin{equation}
	\mu_c(f,T,H) = 1 +  \sum_{m=1}^{M}\frac{\Delta\mu_m(T,H)f_{m}^2(T,H)}{f_{m}^2(T,H)-f^2 - if\gamma_{m}(T,H)},
	\label{mu}
\end{equation}
where, for the $m$-th magnetic resonance, $f_m$ is its frequency, $\gamma_m$ its width and $\Delta\mu_m$ is its input to the zero-frequency magnetic susceptibility. In the case of DFO, $M=2$, with $m=1$ corresponding to the qFMR and $m=2$ to the qAFMR. 

Ferrites have dielectric functions $\epsilon_c(f,T)$, which, 
in our experimental frequency range, i.e.\ far from phonons resonant frequencies, can be approximated as linear around a convenient point ($T_0=400$~K, $f_0=0.3$~THz)
\begin{equation}
\epsilon_c(f,T) = \epsilon_{00} + a(f-f_0) + b(T-T_0).
\label{eps-linear}
\end{equation}

For our polycrystalline samples, we assume an effective\cite{Birchak74, Heller45} magnetic susceptibility as $\sqrt{\mu}=p\sqrt{\mu_c}+(1-p)$ and effective dielectric function as $\sqrt{\epsilon}=p\sqrt{\epsilon_c}+(1-p)$, where the factor $p=0.64$ is a volume fraction of material to air in our pelletized samples.\cite{Bialek19} This value is the maximum density of random-packed hard spheres.\cite{Song08} This assumption allows us to obtain values of $\epsilon_c$ and $\mu_c$ that are comparable with literature values for single crystal samples.\cite{Kozlov93, Lobo07}
The complex wave vector is $k = 2\pi f \sqrt{\epsilon\mu}/c$, where $c$ is the speed of light in vacuum. Transmission of electric field $t(f,T,H)$ though a slab of a thickness $d$ and infinite lateral dimensions is: \cite{BORN}
\begin{equation}
 t(f,T,H) =  \frac{(1-r^2)e^{ikd}}{1-r^2e^{i2kd}},
 \label{slab}
\end{equation}
where $r^2=(\sqrt{\epsilon}-\sqrt{\mu})^2/(\sqrt{\epsilon}+\sqrt{\mu})^2$ is the square of the reflection coefficient at the vacuum-material interface.
For frequencies far away from resonances, Eq.\ \ref{slab} implies an interference pattern, related to subsequent cavity modes of a slab, called Fabry-P\'erot modes. At 300~GHz, in our polycrystalline DFO samples with the refractive index $\sqrt{\epsilon\mu}\approx3.6$, the wavelength is about 300~$\mu$m. With rising temperature, $\sqrt{\epsilon\mu}$ increases, which shortens the period of the interference pattern. The magnetic susceptibility takes into account the effect of the magnetic resonance. When its frequency is close to a mode of a cavity, both the magnon and the cavity lines are altered by the matter-photon interaction. Since, the magnetic resonance frequencies have a much stronger temperature dependence than that of the interference pattern, they cross several interference pattern minima as temperature rises. We calculated
\begin{equation}
\left\|\frac{\partial S_{21}}{\partial T}\right\| = \Delta T^{-1}20log_{10}\left\|\frac{t(f,T+\Delta T,0)}{t(f,T,0)}\right\|
\label{dS21dT}
\end{equation}
to fit amplitude of temperature-differential spectra, 
where $\Delta T=1$~K is the temperature step. 
We calculated phase of temperature-differential spectra using
\begin{equation}
\arg(\partial S_{21}/\partial T) = \arg(t(f,T+\Delta T,0)-t(f,T,0)).
\end{equation}

\begin{figure*}
	\includegraphics[width=0.99\linewidth]{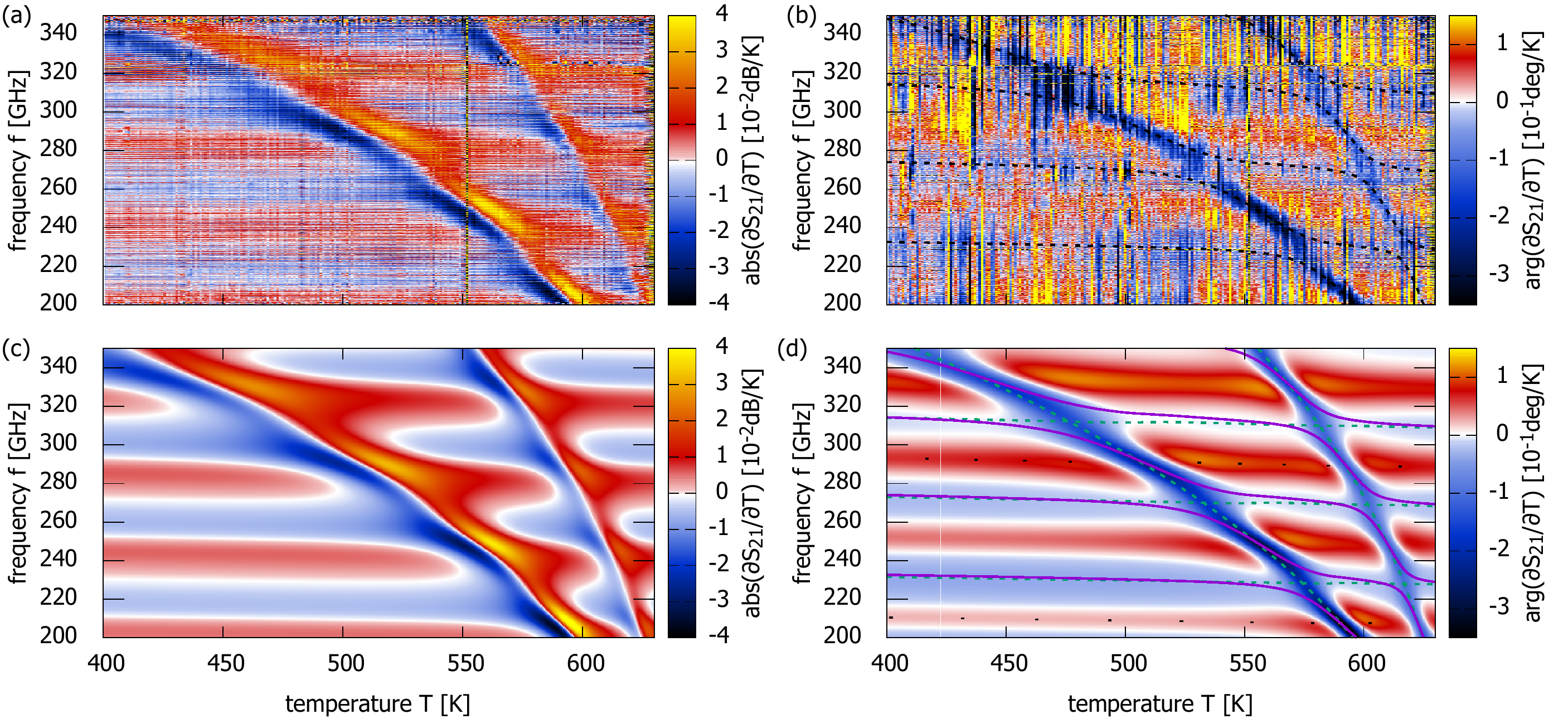}
	\caption{Experimental temperature-differential spectra for the 1.0~mm-thick DFO sample: (a) amplitude, (b) median-shifted phase with dashed black lines showing coupled modes. Fit of the model to the amplitude data: (c) magnitude, (d) phase. In the segment (d), green dashed lines show uncoupled spin-wave and cavity modes and purple lines show coupled modes.}
	\label{1000um-dT}
\end{figure*}

The experimental temperature-differential signal amplitude for the 1.0-mm-thick DFO sample is shown in Fig.\ \ref{1000um-dT}a.
These data show clearly that the resonant lines are distorted when they cross the sequence of sample-cavity modes. The amplitude and the widths of the resonances are altered because of their interaction with the electromagnetic standing waves. This is accounted for by Eq.\ \ref{dS21dT} as shown in Fig.\ \ref{1000um-dT}c. Thus, we find that this model reproduces most of the important features of Fig.\ \ref{1000um-dT}a. The fitting parameters are parameters of the simplified dielectric function (Eq.\ \ref{eps-linear}) and of the magnetic susceptibility (Eq.\ \ref{mu}). We assumed that resonance frequencies have a temperature dependence described by a power law $f_{m}^*(1-T/T_N)^{\beta_m}$, applicable when approaching the N\'eel temperature $T_N$,\cite{Eibschutz66} 
where $f_m^*$ has a unit of frequency and $\beta_m\approx\frac{1}{3}$ in the case of DFO. 
To improve the quality of our fits, we assumed that magnetic resonances have linear dependences of their widths and amplitudes on temperature. These formulas and values of fit parameters are given in the supplementary materials. 

\begin{figure}
	\includegraphics[width=0.99\linewidth]{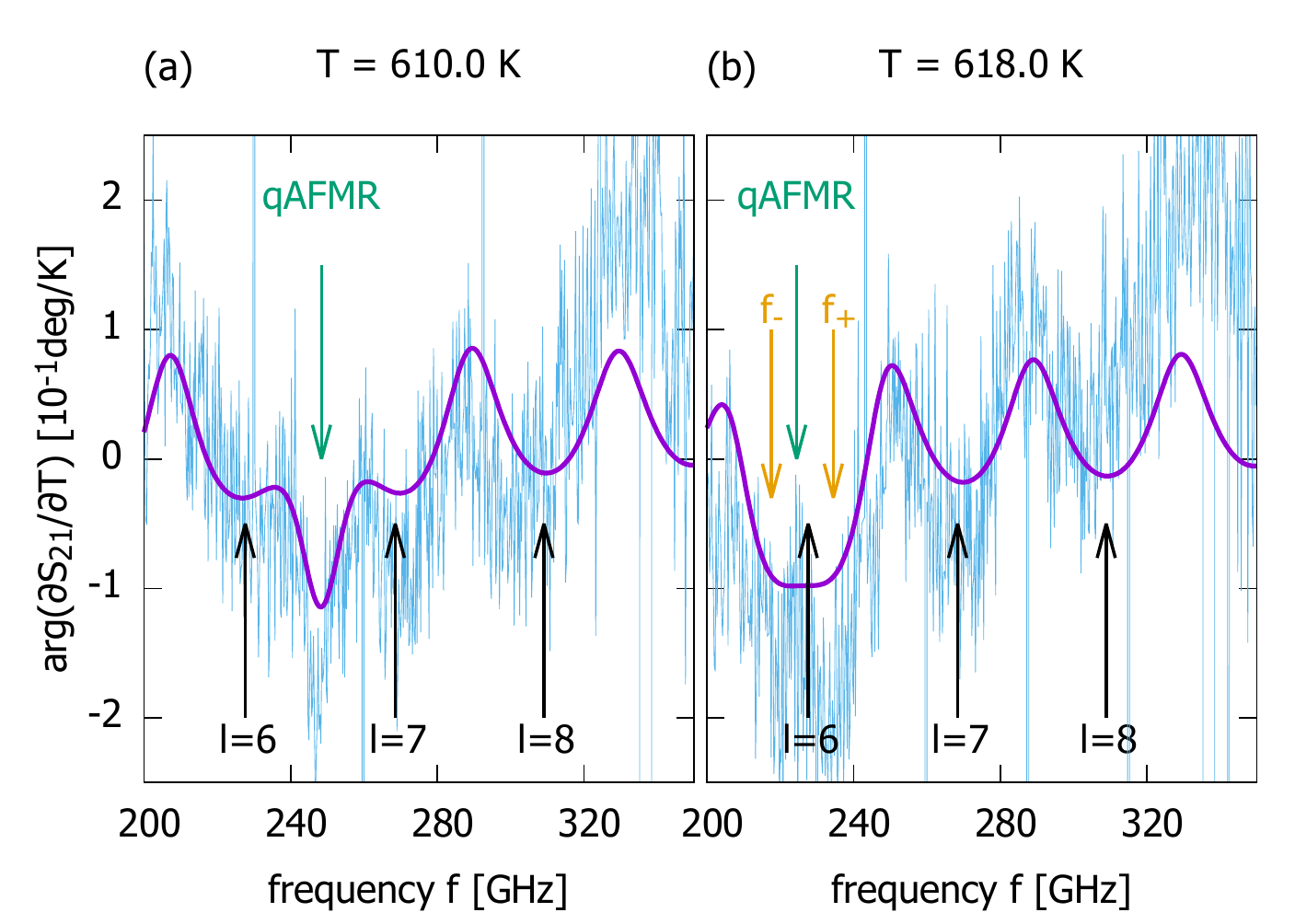}
	\caption{Examples of phase spectra extracted from Fig.\ \ref{1000um-dT}b. (a) spectrum at a temperature not showing light-matter interaction, (b) temperature close to a crossing point. Arrows mark positions of cavity modes, qAFMR and upper and lower polaritons. The violet lines show fits obtained using data in the entire temperature range (Fig.\ \ref{1000um-dT}d), thus better reflecting average properties of the sample.}
	\label{single-spectra}
\end{figure}

We measured the temperature-differential phase of transmitted electric field, as presented in Fig.\ \ref{1000um-dT}b. Despite high noise, features predicted in Fig.\ \ref{1000um-dT}d are observed in the experiment.
The phase reveals clearly interactions between electromagnetic waves and magnetization dynamics. In Fig.\ \ref{single-spectra}a, a spectrum obtained at a temperature between crossings points shows that the qAFMR can be accounted for with a harmonic model. In Fig.\ \ref{single-spectra}b, the spectrum obtained at the temperature of the crossing with the $l=6$ cavity mode shows a structure that cannot be explained using a single oscillator. It is due to two polariton states which are not well-separated, i.e. they are in a weak coupling regime.\cite{Khitrova06} 
We used this phase prediction to estimate the cavity mode-magnetic resonance coupling strength. Thus, we superimposed on the Fig.\ \ref{1000um-dT}d predictions of the harmonic coupling model \cite{Mills74, Huebl13, Harder16}
\begin{equation}
f_{\pm} = \frac{1}{2}\left(f_{(l)} + f_m \pm\sqrt{(f_{(l)}-f_m)^2+4\kappa^2f_{(l)}}\right),
\label{anticross}
\end{equation}
where $f_{\pm}$ indicates upper and lower polariton frequencies, $f_{(l)}$ is the $l$-th cavity mode frequency, $f_m$ with $m=1,2$ are the resonances frequencies. The coupling strength is given by \cite{Huebl13}
\begin{equation}
\kappa = \frac{g_s\mu_B}{2h}\sqrt{\frac{\mu_0h}{2}p\rho},
\label{kappa}
\end{equation}
where $g_s=2$, $\rho$ is density of resonators and $p=0.64$ is the mass filling factor of our polycrystalline sample. \cite{Song08, Bialek19} Equation \ref{kappa} assumes that a magnon is coupled to a single electromagnetic cavity mode with a coupling strength\cite{Soykal10PRL} $g_s\mu_BB_0/2h$, where $B_0 = \sqrt{\mu_0h f_{(l)}/2V_{(l)}}$ is the magnetic component of vacuum fluctuations.\cite{Niemczyk09} Amplitude of these is only $\approx 5\cdot10^{-11}$~T for an electromagnetic mode at 0.3~THz taking the volume $V_{(l)}\approx16\pi10^{-9}$~m$^3$ of the 1-mm-thick sample. This is about 3 orders of magnitude smaller than the amplitude of the THz field in our experiment. However, in an ensemble of $N$ resonators that collectively interact with a cavity mode, coupling is increased by a factor $\sqrt{N}$. \cite{Raizen89, Soykal10PRL, Huebl13, Harder16, Li18} Thus, the collective coupling strength $\kappa$ depends only on oscillators density $\rho=N/V_c$ and the ratio $p$ of a crystalline volume $V_c$ to a cavity volume $V_{(l)}$ \cite{Huebl13, Li18}. We take $\rho=\frac{1}{3}\rho_{Fe}$, where $\rho_{Fe}=1.76\times10^{28}$~m$^{-3}$ is density of iron atoms in DFO.\cite{Eibschutz65} This factor reflects the fact that in a polycrystalline material, on average, only $1/3$ of magnetic dipoles are excited by linear polarized electromagnetic wave. Under this assumption $\kappa=1.78\times10^4$~Hz$^{1/2}$, which results in a splitting $2\kappa \sqrt{f_{(l)}}\approx 19.5$~GHz with a cavity mode of $f_{(l)}=0.3$~THz.


Using Eq.\ \ref{anticross}, we calculated modes undergoing subsequent interactions with the same coupling strength $\kappa$. 
We determined $f_{(l)}$ from the condition $c(l-1/2)=2\Re{(\sqrt{\epsilon\mu})}f_{(l)}d$ for minimum of $\arg(\partial S_{21}/\partial T)$.
In the case of 1-mm-thick sample, the lowest visible mode at $\approx 230$~GHz has $l=6$. Resonant modes frequencies were obtained from the same fit to the experimental $\partial S_{21}/\partial T$ magnitude. For the 1.0-mm-thick sample, the temperature dependence of both coupled and uncoupled modes is presented in Fig.\ \ref{1000um-dT}d.

In summary, we have observed coupling of spin-waves with electromagnetic fields in high-temperature antiferromagnet DyFeO$_3$. We observed the coupling of THz-frequency magnetic resonances to modes of cavities formed by the samples themselves. 
Our research demonstrates a possibility for a spin-cavitronic system based on antiferrromagnetic resonances. To account for our results, we applied both classical electrodynamics and a microscopic magnon-photon coupling model,\cite{Huebl13} taking into account the polycrystalline nature of our samples.

\begin{acknowledgments}
We would like to thank prof. Can-Ming Hu of University of Manitoba for very fruitful discussions and Claude Amendola for help in the samples preparation.
This work was partially supported by a Requip (206021150707/1) grant of the Swiss National Science Foundation and by EPFL.
\end{acknowledgments}

\bibliography{polaritonsDFOBFO}

\end{document}